# Multiscale Approach for Bone Remodeling Simulation Based on Finite Element and Neural Network Computation

**Ridha HAMBLI,** Abdelwahed BARKAOUI
PRISME Institute, EA4229, University of Orleans
Polytech' Orléans, 8, Rue Léonard de Vinci 45072 Orléans, France
ridha.hambli@univ-orleans.fr

## ABSTRACT

In this paper, a novel multiscale hierarchical model based on finite element analysis and neural network computation was developed to link mesoscopic and macroscopic scales to simulate bone remodeling process. The finite element calculation is performed at macroscopic level and trained neural networks are employed as numerical devices for substituting the finite element computation needed for the mesoscale prediction. Based on a set of mesoscale simulations of representative volume elements of bone taken from different bone sites, a neural network is trained to approximate the responses at meso level and transferred at macro level.

## 1. Introduction

The bone adaptation process called remodeling process must address changes in its morphological and mechanical properties over time, at multiple levels, allowing for a more accurate description of the bone architecture [1-2]. This process occurs at different time and spatial scales in hierarchical way with interacting phenomenon between the different scales [1-2]. Therefore, an accurate simulation of bone remodeling should include different length scales. In the last few years, a general strategy has emerged for the design of multiscale methods in order to capture the macroscale behavior of the solutions. A large number of methods are based on numerical homogenization procedure [3], others introduce a new methods, and among them the artificial neural networks (NN) has attracted a growing interest in recent years [4]. In this paper, a hybrid method for bone remodeling multiscale simulation using finite element analysis and NN computation is proposed. The motivation to develop such hierarchical approach in the frame work of bone modeling is that a part of the mechanisms affecting bone strength are observed at trabecular level and cannot be described in precise way at macro level like microcracks accumulation, stress concentration du to tabecular network, bone cells activities and mass or chemical transport. Meshing the entire femur with its trabecular architecture generates some millions of finite elements with a huge computational time. The NN approach is beneficial if the numerical analysis of the complex model is time consuming or unfeasible [4].

## 2. Method

The hybrid FENN method is a simulation procedure in which a continuum model is discretized into smaller submodels composed of RVE (fig. 1). Changes in the material distribution at macro level will have an effect on the stress/strain field, thus affecting the mechanical state of each RVE in the subsequent iteration. At the completion of every iteration, a new FE analysis is performed to update the mechanical parameters distribution in the macro level. A trained NN is applied locally to determine structural and mechanical change at meso level. The local results are passed back to the macro level.

The proposed methodology follows this iterative procedure until convergence is achieved. A schematic illustration of the hierarchical approach is presented in Fig.1.

The meso approach can be summarized by the five following steps:
(i) FE remodeling simulation of the RVE for different combinations of bone inputs
(ii) Averaging the RVE outputs.
(iii) Steps (i) and (ii) supply training data in the form of a 'design of experiments' (DoE) table for NN training.
(iv) Training the NN based on the numerical DoE.
(v) Incorporation of the NN into the macro FE model to link meso-to-macro scales.

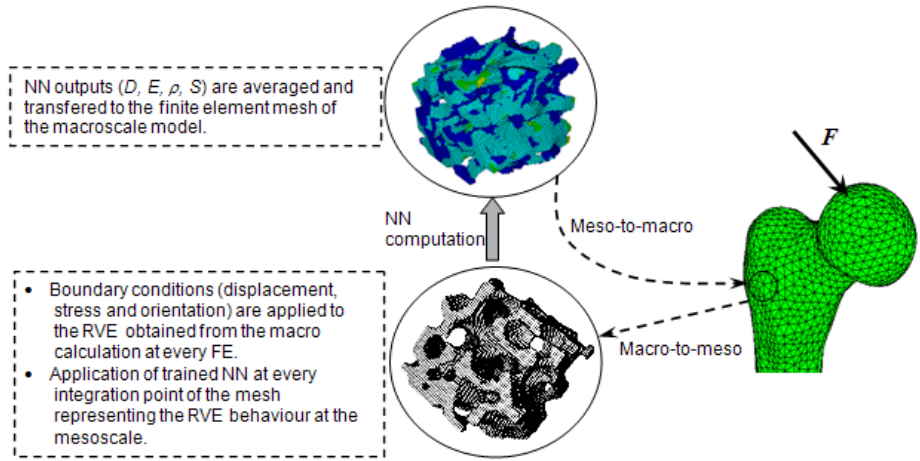

**Figure 1.** Multiscale hierarchical approach for bone analysis. NN is incorporated into the FE code Abaqus via the routine UMAT.

## 3. Neural network

Neural Networks models are composed of a large number of inter-connected processing elements called neurons, organized in layers. [8]. The single neuron performs a weighted sum of the inputs $x_i$ that are generally the outputs of the neurons of the previous layer $v_m$, adding threshold value $b_i$ and producing an output given by:

$$v_m = \sum_{i=1}^{L} w_{im} x_i + b_i \qquad (1)$$

$w_{im}$ are the network weights.

Input signals cumulated in the neuron block are activated by a nonlinear function given by:

$$f(v_m) = \frac{1}{1 + exp(-\beta v_m)} \qquad (2)$$

Where $\beta$ is a parameter defining the slope of the function.

To apply the NN, a training phase is needed which consists of an optimization procedure in order to determine the weights of the NN.

20 finite element simulations have been performed on every RVE model corresponding to the combinations of local applied stress (4 levels) and cycle frequency (5 levels) covering the range of mechanically observed/computed stresses and fequencies [5, 6].

## 4. Finite element model for bone remodeling

The set of equation describing the fully coupled change in the bone density is given by [5, 6]:

$$\frac{d\rho}{dt} = \alpha_R (S - S_R) \quad \text{If} \quad S < S_R \tag{3}$$

$$\frac{d\rho}{dt} = 0 \quad \text{If} \quad S_R \leq S \leq S_F \tag{4}$$

$$\frac{d\rho}{dt} = \alpha_F (S - S_F) \quad \text{If} \quad S > S_F \tag{5}$$

$$\frac{d\rho}{dt} = \alpha_D (S - S_D) \quad \text{If} \quad S > S_D \tag{6}$$

Where $\rho$ is the bone density, $t$ is time and $S$ is the coupled strain-damage stimulus function. $\alpha_R$, $\alpha_F$ and $\alpha_D$ denote respectively bone resorption rate, bone forming rate and damage resorption rate.
$S_R$, $S_F$ and $S_D$ denote respectively target levels of strain-damage energy density for bone resorption, formation and damage resorption.

The averaging relation of each RVE outputs, $y_i^{RVE}$, is expressed by :

$$y_i^{RVE} = \frac{1}{V_o} \int_{V_o} y_i \, dV \tag{7}$$

Where $V_o$ and $y_i$ denote respectively the RVE reference domain and the output at every finite element location $i$.

## 5. Simulation of 3D femur head remodeling

To illustrate the capabilities of the FENN method, the remodeling of a 3D model of femur head has been studied. There are two materials in the macro model, trabecular bone and cortical bone. Marrow is considered only in the porous regions at the RVEs level. A 3D mesh is generated using tetrahedral elements. The model is run in alternating load and unload increments during 1000 iterations (days) with a fixed number of cycles per day (7000 cycles/day).

Since the investigation scale of the present work corresponds to one or some trabeculae level, we assume that the bone behaviour is purely elastic with isotropic averaged properties from nanoscale level.

## 6. Results

In Figure 2 the distribution of the bone density is presented for both macro (a) and meso levels (b). The local RVE result is averaged and transferred to the entire femur for more accurate prediction of bone density.

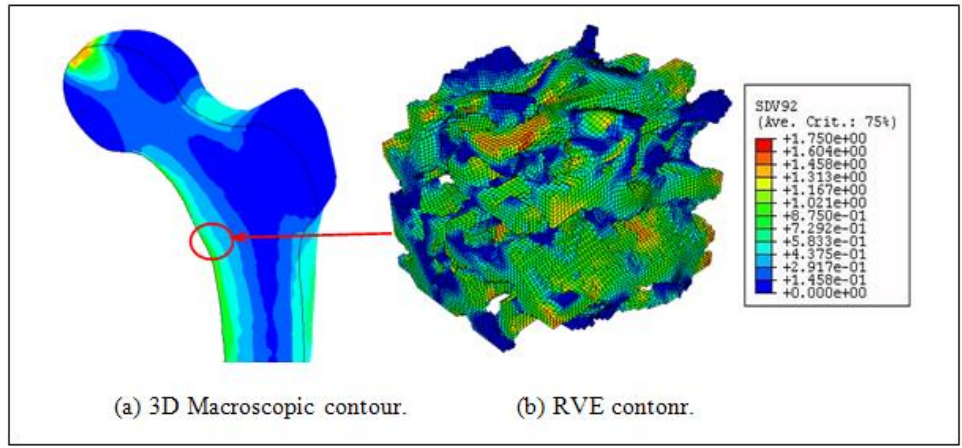

(a) 3D Macroscopic contour.    (b) RVE contonr.

**Figure 2.** Contour of trabecular bone density predicted by the hybrid method. (a) Macroscopic contour predicted by NN computation. (b) RVE predicted results are averaged and passed to the macroscopic finite element level.

**Conclusion**

In this study, the focus was on the development and the implementation of a novel multiscale approach for bone remodeling simulation using finite element simulation and neural network computation. The NN algorithm has been incorporated into a finite element code to link meso and macro scales and provide local and rapid computation at the meso scale. Actually, the proposed multiscale approach doesn't take into account biological mechanisms. The aim was to provide a framework using FE and NN methodology for a rapid and accurate multiscale modeling of bone.

**Acknowledgements**

This work has been supported by French National Research Agency (ANR) through TecSan program (Project MoDos, n°ANR-09-TECS-018).